\documentclass[prl,aps,10pt,superscriptaddress,twocolumn,floatfix,nofootinbib]{revtex4-1}
\usepackage{graphicx,color}
\usepackage{amsmath,amssymb,bm}
\usepackage[plainpages=false,pdfpagelabels,colorlinks=true,linkcolor=red,urlcolor=blue,citecolor=blue,
pdftitle={Quantum gravity and the spin of black holes in gravitational wave events},
pdfauthor={Eugenio Bianchi, Anuradha Gupta, Hal M. Haggard, B. S. Sathyaprakash},
pdfdisplaydoctitle=true,pdfduplex=DuplexFlipLongEdge,bookmarks=false]{hyperref}

\usepackage[normalem]{ulem}

\begin{document}

\title{Small Spins of Primordial Black Holes from Random Geometries:\\[0.25em]
 Bekenstein-Hawking Entropy and Gravitational Wave Observations}

\author{Eugenio Bianchi}
\affiliation{Department of Physics, The Pennsylvania State University, University Park, PA 16802, USA}
\affiliation{Institute for Gravitation \& the Cosmos, The Pennsylvania State University, University Park, PA 16802, USA}
\author{Anuradha Gupta}
\affiliation{Department of Physics, The Pennsylvania State University, University Park, PA 16802, USA}
\affiliation{Institute for Gravitation \& the Cosmos, The Pennsylvania State University, University Park, PA 16802, USA}
\author{Hal M. Haggard}
\affiliation{Physics Program, Bard College, 30 Campus Road, Annandale-On-Hudson, NY 12504, USA}
\affiliation{Perimeter Institute, 31 Caroline Street North, Waterloo, ON, N2L 2Y5, CAN}
\author{B. S. Sathyaprakash}
\affiliation{Department of Physics, The Pennsylvania State University, University Park, PA 16802, USA}
\affiliation{Institute for Gravitation \& the Cosmos, The Pennsylvania State University, University Park, PA 16802, USA}
\affiliation{School of Physics and Astronomy, Cardiff University, Cardiff, CF24 3AA, UK}

\begin{abstract}
Black hole entropy is a robust prediction of quantum gravity with no established phenomenological consequences to date. We use the Bekenstein-Hawking entropy formula and general-relativistic statistical mechanics to determine the probability distribution of random geometries uniformly sampled in phase space. We show that this statistics (in the limit $\hbar\to 0$) is relevant to large curvature perturbations, resulting in a population of primordial black holes with zero natal spin. In principle, the identification of such a population at LIGO, Virgo, and future gravitational wave observatories could provide the first observational evidence for the statistical nature of black hole entropy. 
\end{abstract}

\maketitle

\emph{Black holes: simple or complex}. Black holes are often considered to be the `simplest' macroscopic gravitating objects. They are vacuum solutions of General Relativity that, at equilibrium, are fully characterized by their mass and spin \cite{Bardeen:1973gs}. In 1974, however, Hawking discovered that black holes are hot because of quantum effects neglected in Einstein's classical theory of gravity. As a result, black holes have an entropy---a distinguishing feature of complex systems like hot gases. In fact, black holes have a huge entropy, much larger than the entropy of a star of the same mass. This entropy, given by the Bekenstein-Hawking formula \cite{Bekenstein:1973ur,Hawking:1974sw}, is proportional to the area $A$ of the black hole horizon, and for a rotating black hole of mass $M$ is given by the equation
\begin{equation}
S(M, a)=\frac{A(M, a)}{4\ell_P^2}=\big(1+\sqrt{1- a^2}\, \big)\,\frac{2\pi M^{2}}{m_{P}^{2}}\,,
\label{Ent}
\end{equation}
where $\ell_P=\sqrt{\hbar G/c^{3}}\,$ is the Planck length, $m_{P}=\sqrt{\hbar c/G\,}$ is the Planck mass, and 
\begin{equation}
 a=\frac{J}{G M^2/c}\;\in [0,1] \,,
\label{eq:}
\end{equation}
is the dimensionless spin parameter. Considerable effort has been devoted to deriving this formula from the microscopic degrees of freedom of prospective theories of quantum gravity. Despite differing approaches, the result is found to be robust: In the interaction with its surroundings, a black hole of mass $M$ and spin $ a$ behaves as an ensemble consisting of $\mathcal{N}\sim e^{\,S(M, a)}$ microstates. In this letter we investigate a phenomenological implication of this prediction shared by all current approaches to quantum gravity: we describe how the statistical nature of black holes typically results in zero spin.

\begin{figure}[t!] 
\includegraphics[width=0.47\textwidth]{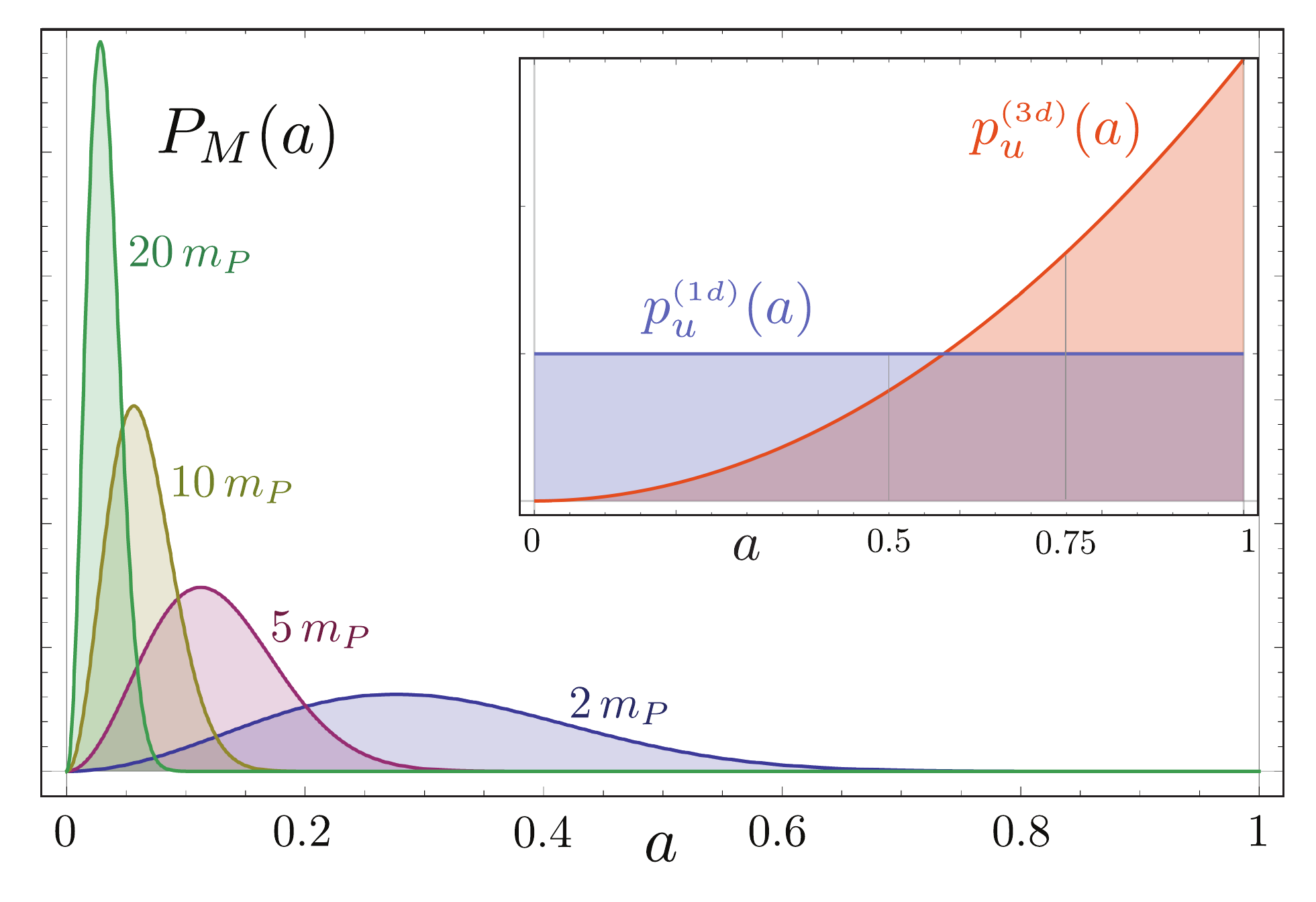}   
\caption{Probability distribution of spins $ a$ for black holes of mass $M$ in the microcanonical ensemble. At large mass, the average spin is small. The inset contrasts two alternate distributions, one uniform in $a$ and one uniform in $\vec{a}$.}
   \label{fig:Pofchi}
\end{figure}

\medskip

\begin{figure*}[t!] 
${}$\hspace{-11pt}\includegraphics[width=0.97\textwidth]{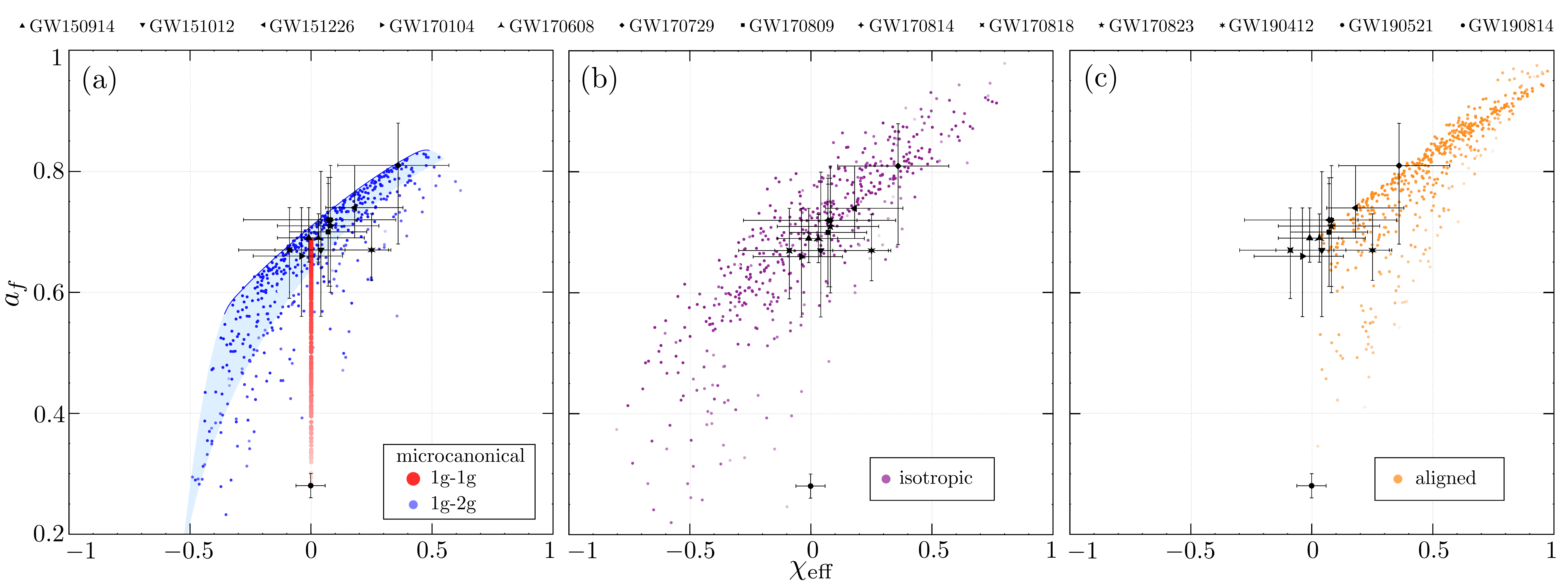}   
\caption{Final spin, $ a_{f}$, versus effective spin parameter, $ \chi_{\text{eff}}$, of binary BH mergers in different scenarios. The binary constituent masses $M_1$ and $M_2$ are sampled uniformly in the range $[10M_{\odot}, 100M_{\odot}]$ so that $M_1/M_2\leq 10$.  Panel (a): 1g-1g population consisting of mergers of two BHs drawn from the \emph{microcanonical} ensemble with spin distribution given by Eq.~(\ref{eq:Pofchi}) [red dots], and 1g-2g population consisting of mergers of a 1g BH with the product of a 1g-1g merger [blue dots], with the blue shaded region $M_1/M_2 \leq 4$. Panel (b): Mergers of two BHs with isotropically distributed spins. Panel (c): Mergers of two BHs with aligned spins.  The darkest colors in the gradient indicate equal mass, while the lightest colors indicate the most asymmetric binary masses. The black symbols show the GW events measured thus far by the LVC detectors \cite{aLIGO:ref, AdV:ref} and the error bars on their $\chi_{\rm eff}$ and $a_f$ values are $90\%$ credible bounds \cite{Abbott:2016blz,Abbott:2016nmj,TheLIGOScientific:2016pea,Abbott:2017vtc,Abbott:2017oio,Abbott:2017gyy,LIGOScientific:2018mvr,LIGOScientific:2018jsj,LIGOScientific:2020stg,Abbott:2020tfl,Abbott:2020khf}. }
\label{fig:chif_chieff_plot}
\end{figure*}

\emph{Black hole entropy and the spin distribution}. The Bekenstein-Hawking formula shows that, at fixed mass $M$, black holes with larger spin have a smaller entropy. The statistical mechanical interpretation of this formula implies that---at fixed mass---there are fewer microstates with large spin than with small spin \cite{Bianchi:2018uci}. As a result, in the statistical ensemble where only the energy of the system is held fixed (the microcanonical ensemble), the probability of finding spin $ a$ is given by the fraction
\begin{equation}
P_{M}( a)=\frac{\displaystyle e^{A(M, a)/4\ell_{P}^{2}}\, a^{2}}{\displaystyle \int_{0}^{1}\;e^{A(M,a')/4\ell_{P}^{2}}\;{a'}{}^{2}\,da'}\;,
\label{eq:Pofchi}
\end{equation}
where the numerator counts the number $\mathcal{N}$ of microstates at fixed mass and spin, while the denominator is the total number of microstates with fixed mass $M$. A formal derivation of this probability distribution from \emph{random geometries and general-relativistic statistical mechanics} is provided below, together with a discussion of its relevance to primordial black holes. Fig.~\ref{fig:Pofchi} shows the distribution $P_{M}( a)$ for a variety of masses, and illustrates its concentration at zero spin for large mass. For a population of black holes distributed according to the microcanonical ensemble, the Bekenstein-Hawking formula implies that, for mass $M\gg m_{P}$, the average spin is small: $\langle  a\rangle\approx 2m_{P}/\pi M \ll 1$. For a solar-mass black hole, this is an angular momentum of order $10^3\!$ J$ \cdot$s, corresponding to a dimensionless spin parameter $\langle a\rangle\approx 10^{-38}$. 
This prediction of the Bekenstein-Hawking formula and the existence of black holes (BH) with spin $a \ll 0.7$ can be tested through gravitational wave (GW) observations of black hole spins.

\medskip

\emph{Observation of black hole spins in GW events}. A binary BH merger can be modeled as a process with \emph{in} and \emph{out} data given by
\begin{equation}
(M_{1},\vec{ a}_{1})\;+\;(M_{2},\vec{ a}_{2})\;+\;\vec{L}\;\longrightarrow\; (M_{f},\vec{ a}_{f})\;+\;GW\,,
\label{eq:}
\end{equation}
where $(M_{i},\vec{ a}_{i})$ with $i=1,2\,$ is the mass and dimensionless spin of each BH in the binary, $\vec{L}$ their initial orbital angular momentum, \emph{GW} the emitted gravitational waves in the merger, and $(M_{f},\vec{ a}_{f})$ the mass and spin of the final BH. Gravitational wave observations can be used to measure the final spin magnitude $ a_{f}=|\vec{ a}_{f}|\in [0,1]$ and the orbital projection of the effective spin of the BH binary,
\begin{equation}
 \chi_{\text{eff}}=\frac{M_{1}\,\vec{ a}_{1}+M_{2}\,\vec{ a}_{2}}{M_{1}+M_{2}}\cdot\frac{\vec{L}}{|\vec{L}|}\;\;\in\; [-1,+1]\,,
\label{eq:}
\end{equation}
a quantity conserved at the 2nd post-Newtonian order \cite{Blanchet:2013haa}. Current observations of GW events indicate that the effective spin $ \chi_{\text{eff}}$ of the few BH binaries whose mergers have been observed so far is small and compatible with zero \cite{Abbott:2016blz,Abbott:2016nmj,TheLIGOScientific:2016pea,Abbott:2017vtc,Abbott:2017oio,Abbott:2017gyy,LIGOScientific:2018mvr,LIGOScientific:2018jsj,LIGOScientific:2020stg,Abbott:2020tfl,Abbott:2020khf}. Unless the spins $\vec{ a}_{1}$ and $\vec{ a}_{2}$ are anti-aligned or lie in the plane of the orbit, a small $ \chi_{\text{eff}}$ indicates small spin magnitudes for the progenitor BHs in the binary. This observation is to be contrasted to the measurement of BHs in $X$-ray binaries which are found to have high spin magnitude \cite{Miller:2014aaa}. In Fig.~\ref{fig:chif_chieff_plot} we report the effective initial spin $ \chi_{\text{eff}}\,$---the most easily accessed spin parameter in GWs---and the final spin $ a_f$ of the thirteen binary BH mergers observed thus far, as reported in the LIGO Scientific and Virgo Collaboration (LVC) catalog \cite{Abbott:2016blz,Abbott:2016nmj,TheLIGOScientific:2016pea,Abbott:2017vtc,Abbott:2017oio,Abbott:2017gyy,LIGOScientific:2018mvr,LIGOScientific:2018jsj,LIGOScientific:2020stg,Abbott:2020tfl,Abbott:2020khf}.

\medskip

\emph{Microcanonical ensemble and GW events}. Gravitational wave observations provide a way to test the microcanonical ensemble determined by the Bekenstein-Hawking entropy. If the progenitor BHs are drawn from the microcanonical ensemble,  then the spin of each BH is distributed according to the probability distribution (\ref{eq:Pofchi}). For BHs of a solar mass or more, this implies that their spin is zero for all practical purposes and $ \chi_{\text{eff}}\simeq 0$. This observation has an immediate consequence for the spin of the final BH. It is known from numerical-relativity simulations that the merger of two zero-spin BHs results in a final BH with spin $ a_{f}$  largely independent from the details of the process: a fit of numerical simulations  provides a phenomenological formula for the final spin \cite{Scheel:2008rj,Hofmann:2016yih}, which truncated to lowest order in the fractional mass difference $\delta=|M_{1}-M_{2}|/(M_{1}+M_{2})\ll1$, is
\begin{equation}
 a_{f}\;\simeq\; 0.686\,-\,0.565\,\delta^{2}\,. 
\label{eq:chifit}
\end{equation}
 Remarkably, for quasi-circular orbits the final spin does not depend on the initial orbital angular momentum $\vec{L}$ of the binary. This feature provides a handle to identify a population of progenitor microcanonical BHs via GW observations of merger events. 

In Fig.~\ref{fig:chif_chieff_plot}(a) we present the distribution of spins predicted by the microcanonical ensemble. We denote by 1g a first generation BH belonging to the microcanonical ensemble. Depending on the formation mechanism, hierarchical mergers are also possible \cite{ClesseGarciaBellido:2015,ClesseGarciaBellido:2017,Fishbach:2017dwv, Gerosa:2017kvu} and we denote by 2g the result of a 1g-1g merger, i.e., a 2nd generation BH. Note that the spin of a 2g BH is isotropically distributed and in general not aligned to the orbital angular momentum of any subsequent 1g-2g merger. For 1g BHs, we consider a uniform distribution of masses in $[10M_{\odot}, 100M_{\odot}]$, where $M_{\odot}\gg m_{P}$ is a reference scale which drops out of the prediction of the spin.  
The spin and mass of the final BH produced from the merger are computed by taking the average of estimates from various fits to numerical-relativity simulations as done for LVC binary BH mergers in the second observing run \cite{Hofmann:2016yih, Jimenez-Forteza:2016oae, Healy:2016lce, Nathan2016}.
Fig.~\ref{fig:chif_chieff_plot}(a) shows a distribution of 1g-1g mergers extracted from this distribution (red dots with $ \chi_{\text{eff}}= 0$ and $ a_{f}\in [0.278,0.686]$). We consider also 1g-2g mergers, with the 2g BH extracted from the previous 1g-1g mergers \cite{Gerosa:2017kvu}. Fig.~\ref{fig:chif_chieff_plot}(a) shows a distribution of 1g-2g mergers (blue dots); notice the dashed line representing the sharp edge of the distribution, given by mass ratios of order $1$.

\medskip

We compare the predictions of the microcanonical ensemble to two astrophysical models of BH spin distributions \cite{Farr:2017uvj,Belczynski:2017gds,Rodriguez:2016vmx}. The first model assumes an \emph{isotropic} distribution of the spins $\vec{ a}_{1}$ and $\vec{ a}_{2}$. Binaries formed in globular clusters or stellar clusters near active galactic nuclei are expected to have isotropic spins \cite{Benacquista:2011kv}. The second model assumes BH spins \emph{aligned} to the orbital angular momentum, i.e., $\vec{ a}_{1}$, $\vec{ a}_{2}$, and $\vec{L}$ pointing in the same direction. Black hole binaries formed through common  envelope evolution in galactic fields are expected to be well-described by this configuration \cite{Postnov:2014tza}. 

While the two models prescribe the directions of the spins, they do not fix their magnitudes. For the isotropic model we assume a uniform distribution for a $3d$ vector, $p_{u}^{(3d)}( a)=3\, a^{2}$, while for the aligned model we assume the distribution $p_{u}^{(1d)}( a)=1$ uniform in the interval $[0,1]$, see inset of Fig.~\ref{fig:Pofchi}. The predictions of the isotropic and aligned models are reported in Fig.~\ref{fig:chif_chieff_plot}(b) and Fig.~\ref{fig:chif_chieff_plot}(c). The Supplemental Material \cite{SupMat}  analyzes the distinguishability of mixed populations. The dashed sharp boundary of Fig.~\ref{fig:chif_chieff_plot}(a) could help distinguish different populations of BHs.

We now show how zero spin arises as the typical configuration in general-relativistic statistical mechanics.

\medskip

\emph{Quantum gravity, the microcanonical ensemble,  and Boltzmann statistics}. The spin distribution (\ref{eq:Pofchi})  relies on the statistical interpretation of the Bekenstein-Hawking entropy in terms of gravitational microstates. While identifying the nature of these microstates requires a theory of quantum gravity (see \cite{Rovelli:1996dv,Ashtekar:1997yu,Engle:2009vc,Bianchi:2010qd,Perez:2017cmj} for results in loop quantum gravity, \cite{Strominger:1996sh,Mathur:2005zp,Mandal:2010cj,Maldacena:2013xja,Harlow:2014yka} for results in string theory, and \cite{Carlip:2008rk} for a discussion of other approaches), counting microstates at fixed mass $M\gg m_{P}$ can be achieved via semiclassical methods \cite{Gibbons:1976ue,Brown:1992bq,Sen:2012dw}. We discuss the derivation of the counting in detail as it clarifies the physics of the microcanonical ensemble \cite{Bianchi:2018uci}. Microstates with asymptotically flat boundary conditions are simultaneous eigenstates of the Arnowitt-Deser-Misner (ADM) mass and spin. They are labeled by their mass $M$, their spin $J=\sqrt{j(j+1)}\,\hbar$ and by a label $\alpha$ that enumerates an orthonormal basis $|M,j,\alpha\rangle$ of the Hilbert space $\mathcal{H}_{Mj}$ at fixed mass and spin. In the microcanonical ensemble, the microstates of given energy $M$ are uniformly populated resulting in a maximally-mixed state
\begin{equation}
\rho_{M}=\frac{1}{\sum_{j'}\text{dim}\, \mathcal{H}_{Mj'}}\sum_{j}\sum_{\alpha}|M,j,\alpha\rangle\langle M,j,\alpha|\,.
\label{eq:deltaH}
\end{equation}
Remarkably, the microcanonical ensemble $\rho_{M}$ consists of a mixture of ensembles $\rho_{Mj}$ of fixed mass and spin, i.e., $\rho_{M}=\sum_{j} p_{M}(j)\,\rho_{Mj}$
where $\rho_{Mj}=\frac{1}{\text{dim}\, \mathcal{H}_{Mj}}\sum_{\alpha}|M,j,\alpha\rangle\langle M,j,\alpha|$ describes the state of a rotating BH. The probability of finding a rotating BH of spin $j$ in the microcanonical ensemble of energy $M$ is therefore given by the ratio
\begin{equation}
p_{M}(j)=\frac{\text{dim}\, \mathcal{H}_{Mj}}{\sum_{j'}\text{dim}\, \mathcal{H}_{Mj'}}\,.
\label{eq:pMj}
\end{equation}
The dimension of the Hilbert space $\text{dim}\, \mathcal{H}_{Mj}$ can be computed via semiclassical methods starting from the canonical partition function $Z(\beta,\omega)=\int \mathcal{D} g_{\mu\nu} \,e^{-I(\beta,\omega)/\hbar}$, where $I(\beta,\omega)$ is the Euclidean gravitational action with fixed periodicity \cite{Gibbons:1976ue}. In the semiclassical limit, taking into account graviton loops in 4d vacuum gravity, one finds
\begin{align}
\text{dim}\, \mathcal{H}_{Mj}\;\sim\;\sqrt{S(M, a_{j})^{\frac{212}{45}-\frac{3}{2}} \,}\;\,e^{\; S(M, a_{j})} \; a_{j}^{2}\,,
\label{eq:dimMj}
\end{align}
and $\sum_{j}\text{dim}\, \mathcal{H}_{Mj}\sim\sqrt{S(M,0)^{\frac{212}{45}} \,}e^{\; S(M,0)}$, where $S(M, a)$ is the Bekenstein-Hawking entropy, Eq.~(\ref{Ent}), and $ a_{j}=\sqrt{j(j+1)}\,\,m_{P}^{2}/M^{2}$ is the dimensionless spin. Therefore, in the limit $M\gg m_{P}$, using \eqref{eq:dimMj} and the probability (\ref{eq:pMj}) reproduces the distribution $P_{M}( a)$ of Eq.~(\ref{eq:Pofchi}). 

We note that considering Standard-Model matter coupled to gravity or considering a different ensemble (for instance including the full mass range $[0,M]$) only has the effect of changing the numerical factor $212/45$ due to gravitons in Eq.~(\ref{eq:dimMj}). As a result, only the logarithmic correction to the Bekenstein-Hawking formula is modified and, in the limit $M\gg m_{P}$,  the probability distribution $P_{M}( a)$ remains unchanged. One can also consider the canonical ensemble at fixed temperature $T$. On time scales much shorter than the black-hole evaporation time, any negative heat capacity instability simply modifies the logarithmic corrections to the Bekenstein-Hawking formula. In this sense, the predictions of the Bekenstein-Hawking entropy and the microcanonical spin distribution $P_{M}( a)$ are robust.

One might assume that quantum statistical mechanics of the gravitational field is only relevant in the deeply non-perturbative or Planckian regime. However, this is not the case. As usual in statistical mechanics, $\hbar$ is required for defining the ensemble, but the expectation value of classical observables, such as the spin, is well defined in the limit $\hbar\to 0$. The classical equations of motion in vacuum are given by the vanishing of the Einstein tensor $G_{\mu\nu}$. In this limit the microcanonical delta function on the Hilbert space, see Eq.~\eqref{eq:deltaH}, becomes a delta function in the covariant classical phase space and we obtain the general-relativistic statistical mechanics of random geometries $g_{\mu \nu}$ described as the uniform Boltzmann distribution at fixed energy $M$, 
\begin{equation}
\label{Boltz}
P_B(g_{\mu\nu}|M) = \frac{ \delta(G_{\mu\nu}) \delta(\mathcal{H}-M) }{\int \mathcal{D} g_{\mu \nu} \delta(G_{\mu\nu}) \delta(\mathcal{H}-M) },
\end{equation}
where $\mathcal{H}$ is the ADM Hamiltonian. The distributional character of Eq.~\eqref{Boltz} means that the well-defined quantities are physical expectation values, given by
\begin{equation}
\!\langle \hat{\mathcal{O}}\rangle_M=\mathrm{Tr}(\hat{\mathcal{O}} \,\rho_M) \stackrel{\hbar\to 0}{\approx} \!\! \int\!\mathcal{D}g_{\mu\nu}\, \mathcal{O}(g_{\mu\nu})\, {P}_B(g_{\mu\nu}| M)\,.
\label{eq:}
\end{equation}
The typical random geometry at fixed mass is predicted to have zero spin
$\langle \hat{J} \rangle  \approx   \int\!\mathcal{D}g_{\mu\nu}\,J\, {P}_B(g_{\mu\nu}| M) = 0$, which agrees with Eq. \eqref{eq:Pofchi} in the limit $\hbar \rightarrow 0$.

\medskip

\emph{Primordial black holes, random geometries, and zero natal spin.}  Primordial black holes (PBH) can be produced by large curvature perturbations generated in the very early universe \cite{Carr:1974nx,GarciaBellido:2017,Sasaki:2018dmp,Carr:2019,Kalaja:2019}. These perturbations freeze as they exit the Hubble horizon, before the end of inflation, and later unfreeze during the radiation era. As a curvature perturbation re-enters it can cause an overdensity throughout a Hubble region of size $H^{-1}$. If the overdensity is above the threshold for gravitational trapping, a BH forms. Such events will, in general, be rare and their frequency will depend on the probability of a large curvature perturbation.  A softening of the equation of state of matter at temperature $T$ results in an exponentially enhanced production of BHs of mass $M \sim H^{-1} \sim m_{P}^3 /T^2$. 

In the formation of PBHs from curvature perturbations, two contributions to the angular momentum must be considered: the one of matter and that of the curvature perturbation. The contribution of thermal matter turns out to be negligible. This is because it scales as $\Delta J \sim \sqrt{V T^3}$ where $V$ is the volume of the region enclosing a given mass. During the QCD transition, $T\approx 150$~MeV gives $M\approx 25\,M_{\odot}$ and  $\Delta J \approx 10^{-4}$ J$\cdot$s.  The second contribution can be predicted from the statistics of  curvature perturbations encoded in the primordial  state $\Psi$ of the geometry. 

Given the state $\Psi$ of the early universe, evolution through an inflationary phase results in a semiclassical probability distribution $P_{\Psi}(g_{\mu \nu})$ for the geometry. In Cosmology, this is generally done by assuming a small quantum curvature perturbation $\zeta(\vec{x}, t)$ over an FRW background, resulting in a nearly gaussian distribution $P(\zeta)$ with an almost scale-invariant power spectrum \cite{LiddleLyth:2000}. This framework is used to describe the temperature anisotropies of the Cosmic Microwave Background. 

Under the assumption of gaussian random fields,  peaks have mild non-sphericity \cite{Bardeen:1986}, and  it has been shown that tidal effects can cause small magnitude spins of order 0.01  \cite{DeLuca:2019} (see also \cite{Chiba:2017rvs, Harada:2017fjm}). Works on the PBH mass spectrum  \cite{Kalaja:2019} acknowledge the potential limitations of the gaussian framework \cite{Franciolini:2018}: ``\dots their mass fraction at formation is extremely sensitive to changes in the tail of the fluctuation distribution and therefore to any possible non-Gaussianity in the density contrast."   These limitations also apply to the prediction of PBH spin from gaussian statistics.

However, to form PBHs, large fluctuations that are rare and strongly interacting are required. These fluctuations go well beyond the two- and three-point correlation functions of the theory and must be drawn from a probability distribution $P_{\Psi}(g_{\mu \nu})$ that is fully relativistic. This probability distribution is diffeomorphism invariant, has FRW as expectation value, and includes background independent features that cannot be captured perturbatively.  

Selecting those random geometries that contain a PBH of mass $M$ is equivalent to conditioning the probability distribution $P_{\Psi}$,
\begin{equation}
\label{PPsiM}
P_\Psi(g_{\mu \nu}|M) = \frac{P_{\Psi}(g_{\mu \nu}) \delta(\mathcal{H}-M)}{\int \mathcal{D} g_{\mu \nu} P_{\Psi}(g_{\mu \nu}) \delta(\mathcal{H}-M)}.
\end{equation}
A complete description of this distribution would require a theory of quantum gravity and the initial condition $\Psi$. Remarkably, under the assumption that $P_\Psi(g_{\mu \nu})$ is sufficiently constant over the mass shell, this distribution reduces to the Boltzmann distribution \eqref{Boltz}. In this way Quantum Mechanics may sample the microcanonical ensemble and produce PBHs with zero natal spin. 

\medskip

\emph{Discussion}. While we have not discussed the PBH mass spectrum here, the full probability distribution $P_{\Psi}(g_{\mu \nu})$ has no obviously preferred window for the production of large perturbations.  A softening of the equation of state of matter exponentially enhances the likelihood of PBH formation from these rare events and can be used to identify the scales at which PBHs are most likely to appear. 

During the QCD transition the pressure drops and the mass of a PBH formed when the temperature is $T\approx 150$~MeV can be estimated to be $M \approx 25 \,M_{\odot}$. Recent studies that take into account results from lattice QCD simulations indicate a range $0.1 - 100\;M_{\odot}$ \cite{Sobrinho:2016fay,Byrnes:2018clq}. This window is directly accessible with the current gravitational-wave observations made by the LVC. 

Other epochs when the effective number of relativistic degrees of freedom drops, such as the electroweak transition and $e^{+}e^{-}$ annihilation, are also accompanied by a softening of the equation of state of matter and could produce PBHs. The mass scale for the first is roughly planetary, and their observation would immediately imply that the BH must be primordial, as we know of no other formation mechanism. The mass scale of the second is $10^6 M_{\odot}$ and observation of a spin much smaller than $0.7$ would indicate that this was a supermassive PBH.

Primordial BHs can form binaries and merge at high redshifts. Observation of correlations of mass, spin, and redshift of binary mergers through GWs can help distinguish subpopulations: the subpopulation that contains astrophysical binary BHs merge at lower redshifts and may or may not have zero effective spins, while a subpopulation of 1g-1g mergers of PBHs can be at high redshifts and have zero effective spins.

\medskip

A developed theory of general-relativistic statistical mechanics does not yet exist (though see \cite{Rovelli:2012}). However, by considering large perturbations in the primordial state of random geometries Eq.~\eqref{PPsiM} we have developed a general-relativistic Boltzmann distribution Eq.~\eqref{Boltz}. Black hole thermodynamics helps us to formulate these Boltzmann statistics, and the simplicity of the Bekenstein-Hawking entropy leads directly to the prediction of zero natal spin from Eq.~\eqref{eq:Pofchi}. We have shown that $J \sim m_{P} M \sim 10^4$ J$\cdot$s for a 25$M_{\odot}$ PBH, or $a \sim 10^{-37}$.

The prediction of small spins can provide new observational constraints on PBHs and their mechanism of formation \cite{Sasaki:2018dmp}.  Most importantly, the imminent transition from single GW measurements to population analyses may provide a new way to investigate the phenomenology of quantum gravity with LIGO, Virgo, and future GW observatories \cite{AmelinoCamelia:2008qg,Barack:2018yly}. In particular, the detection of a population of BHs distributed according to the microcanonical ensemble would provide the first observational evidence for the Bekenstein-Hawking entropy and the statistical mechanics of BHs. 

\medskip

\begin{acknowledgments}
\emph{Acknowledgments}. EB thanks D. Page, B. Unruh, R. Sorkin and V. Frolov for comments and discussions during the 23rd Peyresq meeting, and C. Rovelli for correspondence. We thank N. Johnson-McDaniel for comments on the manuscript. HMH thanks the IGC at the Pennsylvania State University for warm hospitality and support while beginning this work and the Perimeter Institute for Theoretical Physics for generous sabbatical support.  We also thank I. Cholis for correspondence on the formation of primordial binary black holes in galactic halos. This work is supported by Perimeter Institute for Theoretical Physics. Research at Perimeter Institute is supported by the Government of Canada through Industry Canada and by the Province of Ontario through the Ministry of Research and Innovation. EB is supported by the NSF Grant PHY-1806428.  AG and BSS are supported in part by NSF grants PHY-1836779, AST-1716394 and AST-1708146. We acknowledge the use of IUCAA LDG cluster Sarathi for the computational/numerical work. This document has LIGO preprint number {\tt LIGO-P1800228}.
\end{acknowledgments}

\vspace{-1em}

\providecommand{\href}[2]{#2}
\begingroup
\endgroup

\newpage
\phantom{a}
\newpage
\setcounter{figure}{0}
\setcounter{equation}{0}

\renewcommand{\thetable}{S\arabic{table}}
\renewcommand{\thefigure}{S\arabic{figure}}
\renewcommand{\theequation}{S\arabic{equation}}

\renewcommand{\thesection}{S\arabic{section}}

\onecolumngrid

\begin{center}

{\large \bf -- Supplemental Material -- \\[0.25em]
Small Spins of Primordial Black Holes from Random Geometries:\\[0.25em]
 Bekenstein-Hawking Entropy and Gravitational Wave Observations
}\\

\medskip

\medskip

Eugenio Bianchi,$^{1,2}$ Anuradha Gupta,$^{1,2}$ Hal M. Haggard$^{3,4}$, B. S. Sathyaprakash$^{1,2,5}$\\

\medskip

$^1${\it Department of Physics, The Pennsylvania State University, University Park, PA 16802, USA}\\
$^2${\it Institute for Gravitation \& the Cosmos, The Pennsylvania State University, University Park, PA 16802, USA}\\
$^3${\it Physics Program, Bard College, 30 Campus Road, Annandale-On-Hudson, NY 12504, USA}\\
$^4${\it Perimeter Institute, 31 Caroline Street North, Waterloo, ON, N2L 2Y5, CAN}\\
$^5${\it {School of Physics and Astronomy, Cardiff University, Cardiff, CF24 3AA, UK}}

\end{center}

\medskip

\medskip

\twocolumngrid

\label{pagesupp}

\emph{Distinguishing a microcanonical population and a mix of populations}. We compare the predictions of the microcanonical ensemble discussed in \cite{Main} to two astrophysical models of black hole spin distributions \cite{Farr:2017uvj,Belczynski:2017gds,Rodriguez:2016vmx,Postnov:2014tza,Benacquista:2011kv}. The first model assumes that the black hole spins are \emph{aligned} to the orbital angular momentum, i.e., $\vec{ a}_{1}$, $\vec{ a}_{2}$, and $\vec{L}$ pointing in the same direction. The spin magnitudes are assumed to be distributed uniformly, $p_{u}^{(1d)}( a)=1$, according to the flat distribution in the $1d$ interval $[0,1]$. The second model assumes an \emph{isotropic} distribution of the spins $\vec{ a}_{1}$ and $\vec{ a}_{2}$, with magnitudes distributed as $p_{u}^{(3d)}( a)=3\, a^{2}$, the uniform distribution for a vector in $3d$ with magnitude $|\vec{a}|\in [0,1]$. See Fig.~\ref{fig:chif_distribution}.

\medskip

We can use the statistical properties of the $a_{f}$-$\chi_{\rm eff}$ distribution of black hole mergers discussed in \cite{Main} to distinguish a variety of binary black hole populations. We compare the $a_{f}$-$\chi_{\rm eff}$ distribution for an all isotropic population, denoted I, (purple dots in Fig.~2(b) of \cite{Main}), an all aligned one, denoted A,  (orange dots in Fig.~2(c) of \cite{Main}), and various mixed populations, denoted M, using the Anderson-Darling \cite{ADtest} and Kolmogorov-Smirnov \cite{KStest1, KStest2} tests. These tests compare two given distributions and return p-values $\in [0, 1]$. For identical distributions the p-value is 1, and as the difference between the two distributions increases the p-value falls below unity. The mixed populations we consider, M(X,\,Y), all consist of 90\% of population X and 10\% of population Y.  Table~\ref{tab:pvalues} displays the p-value comparisons between pairs of pure and mixed populations. While it is difficult to distinguish a purely isotropic population from an isotropic population with a 10\% admixture of 1g-2g mergers, see the first row of Table \ref{tab:pvalues}, all other mixtures are clearly distinguishable. The improved distinguishability of the 1g-1g over the 1g-2g admixtures is due to the sharp asymmetric peak at $a_{f}\simeq 0.69$ for 1g-1g mergers clearly visible in Fig.~3 of \cite{Main}.

\newpage 

\begin{figure}[t!] 
\setcounter{figure}{2}
\includegraphics[width=0.47\textwidth]{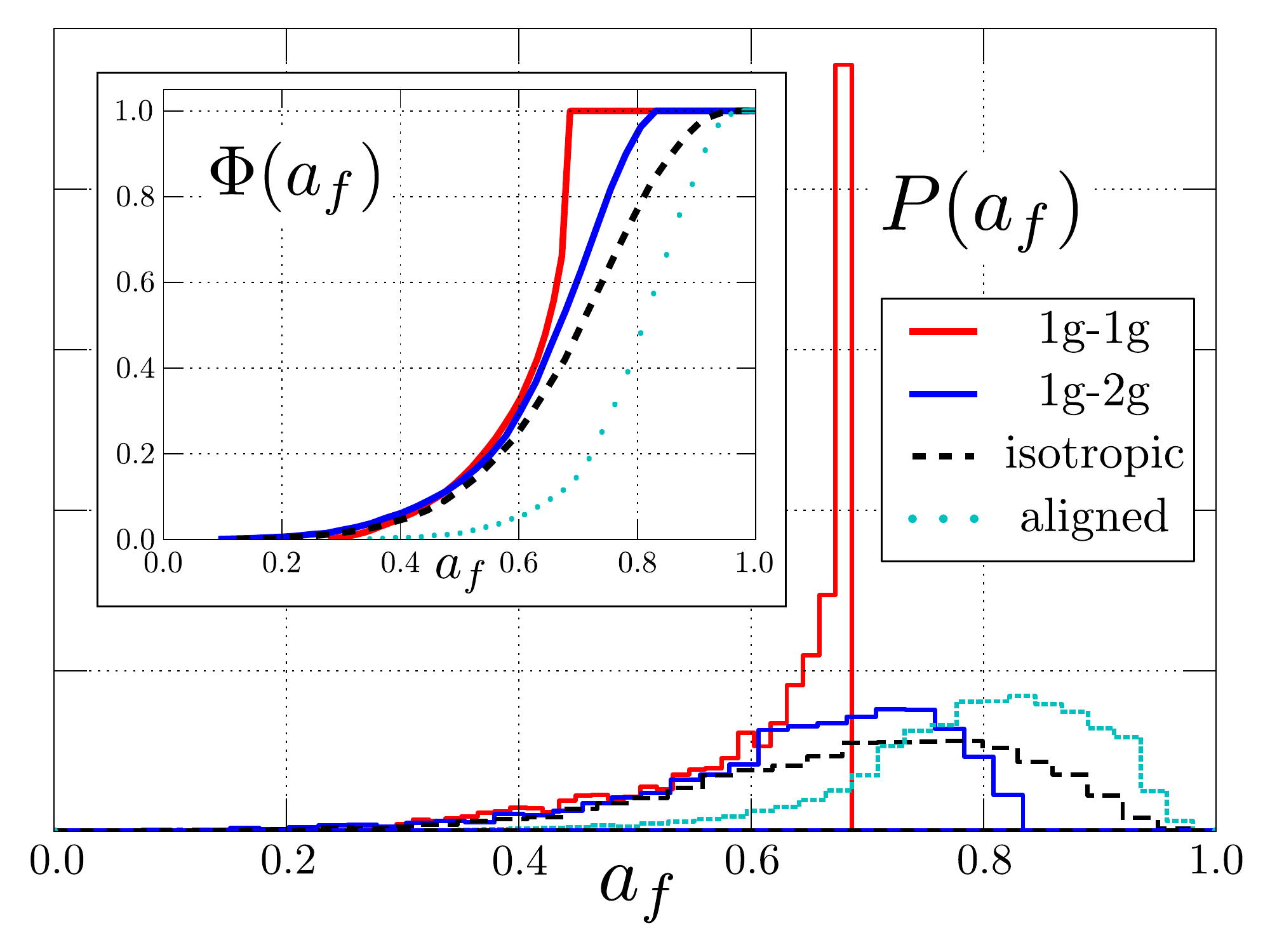}
\caption{Probability distribution of final spins for the models discussed in \cite{Main}. Mergers of 1g black holes with spins distributed according to the probability distribution $P_M(a)$ of \cite{Main} result in a peak at $ a_{f}\lesssim 0.69$. All distributions are normalized to unit probability. \textit{Inset:} Cumulative probability distribution of final spins, $\Phi(a_f)$, for the same models. }
\label{fig:chif_distribution}
\end{figure}

\begin{table}[h]
\begin{center}
\begin{tabular}{|c|c|c|} 
\hline
comparisons \ \ &\   p-value (AD test) \ &\  p-value (KS test)\\
\hline
I vs. M(I,\,1g-2g) & $0.80 $  & $ 0.99  $ \\
I vs. M(I,\,1g-1g) & $  0.03   $  & $  0.02  $\\
A vs. M(A,\,1g-2g)\ \  & $3.1 \times 10^{-5}$ &  $2.8 \times 10^{-3}$ \\
A vs. M(A,\,1g-1g)\ \  & $5.1 \times 10^{-7}$ & $8.7 \times 10^{-7}$\\
\hline
\end{tabular}
\caption{The p-values for Anderson-Darling (AD) and Kolmogorov-Smirnov (KS) tests comparing populations of all aligned (A), all isotropic (I), and mixed (M) binaries. Mixed populations, M(X,\,Y), consist of 90\% of population X and 10\% of population Y.}
\label{tab:pvalues}
\end{center}
\end{table}


\providecommand{\href}[2]{#2}\begingroup
\endgroup

\end{document}